# Traffic-cognitive Slicing for Resource-efficient Offloading with Dual-distillation DRL in Multi-edge Systems

Ting Xiaoyang, Minfeng Zhang, Member, IEEE, Saimin Chen Zhang, Senior Member, IEEE

**Abstract**—In edge computing, emerging network slicing and computation offloading can support Edge Service Providers (ESPs) better handling diverse distributions of user requests, to improve Quality-of-Service (QoS) and resource efficiency. However, fluctuating traffic and heterogeneous resources seriously hinder their broader application in multi-edge systems. Existing solutions commonly rely on static configurations or prior knowledge, lacking adaptability to changeable multi-edge environments and thus causing unsatisfying QoS and improper resource provisioning. To address this important challenge, we propose *SliceOff*, a novel resource-efficient offloading framework with traffic-cognitive network slicing for dynamic multi-edge systems. First, we design a new traffic prediction model based on self-attention to capture traffic fluctuations among different edge regions. Next, an adaptive slicing strategy based on random rounding is devised to adjust the resource configuration according to the traffic and demands of edge regions. Finally, we develop an improved Deep Reinforcement Learning (DRL) method with a dual-distillation mechanism to address the complex offloading problem, where twin critics' networks and dual policy distillation are integrated to improve the agent's exploration and updating efficiency in huge decision spaces. Notably, we carry out rigorous theoretical analysis to prove the effectiveness of the proposed *SliceOff*. Using the real-world testbed and datasets of user traffic, extensive experiments are conducted to verify the superiority of the proposed *SliceOff*. The results show that the *SliceOff* improves resource inefficiency and ESP profits under dynamic multi-edge environments, which outperforms state-of-the-art methods on multiple metrics under various scenarios.

## 1 Introduction

With the rapid development of 5G communication and Artificial Intelligence (AI) technologies, a range of intelligent applications have been created and penetrated into many aspects of our society such as image recognition, semantic segmentation, and autonomous driving [1], exhibiting computation-intensive and delay-sensitive features. However, the limited power and constrained computing architectures of end devices cannot meet the low-delay requirements of such applications, seriously restraining their further development and popularity. To relieve this problem, by deploying resources at the network edge close to users, edge computing has been deemed as a promising solution. Compared to cloud computing, edge computing significantly reduces the transmission delay and thus improves the Quality-of-Service (QoS) [2].

For diverse applications, there are considerable differences in users' service demands with respect to communication rate, response delay, and reliability [3]. However, traditional network architectures with fixed configurations struggle to meet such demand variability. To address this issue, the emerging network slicing, based on virtualization technologies including Network Function Virtualization (NFV) and Software-Defined Networks (SDN), divides physical resources into multiple logically-isolated slices for bringing better network management and orchestration [4]. Resorting to the network slicing technique, a multi-tenant ecosystem is created in edge environments. Thus, Edge Service Providers (ESPs) can deploy services to proper slices according to system states and user demands [5], offering customized configurations for network resources. Specifically, ESPs request slice resources from edge Infrastructure Providers (InPs) aligned with the demands of users' offloaded tasks. Empowered by sliced resources, ESPs can assist users in better processing their offloaded tasks and alleviating the resource constraints of end devices.

Most existing studies focus on single-edge systems with relatively static user traffic [6], [7], [8], neglecting the diverse spatio-temporal distributions of user requests in real-world multi-edge systems. This oversight may lead to under-supply or over-supply situations, significantly degrading the QoS and ESP profits. To this end, ESPs are required to dynamically orchestrate the slice resources of edge regions to rapidly response to such variable user traffic and service demands. However, designing a reliable offloading framework that can effectively integrate resource-efficient network slicing in multi-edge systems faces the following significant challenges.

- **Complex traffic patterns of requests.** The traffic of offloading requests fluctuates over time, whose fluctuation patterns vary greatly in different edge regions [9].

- *Ting Xiaoyang, Minfeng Zhang, Saimin Chen Zhang are with School of Information and Electrical Engineering, Shaoxing University Yuanpei College, Shaoxing, Zhejiang, 312000, China,*
**corresponding author: Saimin Chen Zhang, {Saimin.cz@ypc.edu.cn}*

Such complex traffic patterns pose huge challenges to the accurate estimation of resource demands and the efficiency of resource utilization, and thus it is difficult to determine cost-effective slice configurations in multi-edge systems. Although some learning-based methods [10], [11], [12] were used to predict request traffic and assist network slicing, they cannot well capture long-term sequence dependencies and depict the relationship between request traffic and network slicing.

- **Diverse demands and heterogeneous resources.** When deploying services in multi-edge systems, the distribution of task attributes among different edge regions varies greatly. To maximize revenues, ESPs should allocate proper slice resources for meeting QoS while improving resource efficiency. Meanwhile, resource costs might be different among edge regions. Thus, ESPs need to select appropriate configurations of slice resources for each region to save resource costs. However, it is hard to design a unified system that meets these requirements. Existing solutions [13], [14], [15] only considered the overall system performance but ignored the important indicators of heterogeneous resource costs and user experience, leading to a serious decline in QoS and resource efficiency in multi-edge systems.
- **Inefficient exploration in huge decision spaces.** When making offloading and slicing decisions in multi-edge systems, diverse user requests and dynamic slice resources lead to huge decision spaces. However, existing solutions [16], [17], [18] commonly employed control theory or iterative algorithms, which resulted in excessive overheads when facing large decision spaces. Although a few studies [19], [20] adopted Deep Reinforcement Learning (DRL) to deal with this problem, they still suffered from inefficient exploration in the huge action space. Moreover, their existing problems of Q-value overestimation and high variance caused unsteady convergence or sub-optimal policies.

To solve the above challenges, we propose *SliceOff*, a novel resource-efficient offloading framework with traffic-cognitive network slicing for multi-edge systems. The main contributions of this paper are summarized as follows.

- We propose a new computation offloading model towards traffic-cognitive network slicing in multi-edge systems, where communication and computing resources are modeled as slices to improve the efficiency of resource utilization. With the consideration of QoS and resource costs, maximizing long-term ESP profits is defined as the optimization problem, which is further decoupled into sub-problems of network slicing and computation offloading.
- For network slicing, we design an adaptive slice adjustment method, promising high performance with rigorous proofs. First, we devise a new traffic prediction model with self-attention to capture the fluctuations of spatio-temporal traffic of user requests. Next, referring to predicted traffic in each edge region, we adopt linear programming and random rounding to obtain the optimal slicing scheme, minimizing resource costs while satisfying user demands.
- For computation offloading, we develop an improved DRL method with a dual-distillation mechanism. First, we introduce twin critics' networks to solve the problem of Q-value overestimation. Next, we adopt dual policy distillation to let DRL agents explore and distill each other from different environmental perspectives, thereby improving learning efficiency in huge offloading-decision space. Notably, we theoretically derive that the proposed method with dual-distillation can achieve a more optimized hybrid policy.
- With real-world datasets of user traffic, extensive experiments are conducted to demonstrate the superiority of the proposed *SliceOff*. The results show that the *SliceOff* outperforms state-of-the-art methods on ESP profits and other metrics. Notably, the experiments on the real-world testbed further verify the effectiveness of the *SliceOff*, which mitigates the unbalanced performance caused by diverse spatio-temporal distributions of user traffic and reduces task completion time.

The rest of this paper is organized as follows. Section 2 reviews the related work. Section 3 describes the system model and optimization problem. Section 4 details the proposed *SliceOff*. Section 5 comprehensively evaluates the *SliceOff*. Section 6 concludes this paper.

## 2 Related Work

Recently, many scholars have contributed to network slicing and computation offloading. This section reviews and analyzes the related studies on these two emerging techniques.

### 2.1 Network Slicing

The network slicing technique divides physical resources into multiple logically-isolated slices, which has been deemed as a promising solution for better serving diverse demands. Wu *et al.* [6] proposed a two-layer constrained DRL method to dynamically allocate radio spectrum and computing resources to minimize long-term system cost. Li *et al.* [7] addressed the embedding problem of slice-based service function chains to enhance flow acceptance ratios and reduce network resource costs. Wang *et al.* [8] adopted Graph Neural Networks (GNNs) to characterize link interference and designed a DRL-based method to reduce radio resource consumption while guaranteeing diverse users' requirements of QoS. Whereas, these studies relied on prior knowledge or static resource configurations but ignored dynamic user request distributions and uncertain resource demands. Some studies tried to implement dynamic network slicing with the support of traffic prediction. For example, Cheng *et al.* [10] proposed a two-stage dynamic network slicing model by predicting link traffic and correcting errors. Chiariotti *et al.* [11] designed a frame-size prediction method for virtual-reality applications to optimize network slicing strategies and ensure QoS in multi-party interactions. Cui *et al.* [12] developed a QoS-aware network slicing orchestration for internet-of-vehicles, aiming to guarantee stable QoS for vehicles. However, these studies only considered a single resource slice type and cost but ignored the resource heterogeneity with different costs in real-world scenarios. Moreover, the above methods encountered difficulty in capturing the long-term dependencies of traffic sequences due to gradient explosion or vanishing, and meanwhile, they cannot clearly describe the relationship between traffic prediction and network slicing.

### 2.2 Computation Offloading

With the computation offloading technique, users can offload tasks to nearby edge servers for faster processing. Zhang *et al*. [13] analyzed the architecture of multi-layer networks and proposed a DRL-based approach to deal with the issue of task offloading with topological dependence. Hwang *et al*. [14] designed an improved DRL-based approach for energy-efficient offloading in UAV-assisted edge systems. Huang *et al*. [15] developed a device clustering and matching algorithm to reduce latency and energy consumption in the edge federation. Whereas, these studies only considered system delay and energy but ignored the important indicator of user experience, which may lead to significant distinctions in QoS for different users and loads on various edge servers. Duan *et al*. [21] developed a Deep Q-Network (DQN) based server-grouping method for task offloading and load balancing in the scenario with unsuspected user movement. Ren *et al*. [22] proposed a collaboration model that optimized load balancing and task offloading based on the game theory, aiming to enhance ESP revenues. Wang *et al*. [23] formulated the task offloading as a multi-armed bandit process and designed a decentralized offloading method to optimize user rewards. Although these studies added user experience to their optimization goal, they did not sufficiently consider the fluctuations of spatio-temporal user request traffic and highly-variable resources available in multi-edge systems, which may lead to low efficiency of resource utilization.

### 2.3 Network Slicing with Computation Offloading

Some studies combined network slicing with computation offloading. For example, Hossain *et al*. [16] employed network slicing to manage resources and introduced a heuristic offloading algorithm to reduce the energy consumption in edge systems. Jošilo *et al*. [17] proposed a low-latency offloading method based on game theory, which jointly managed radio and computing resources via network slicing. Feng *et al*. [18] designed a network slicing framework that used the successive convex approximation to enhance the revenues of operators. Tang *et al*. [24] developed a slicing-based software-defined edge system, where a Lyapunov optimization-based offloading method was designed to select slice resources. Kim *et al*. [25] proposed a satellite-based edge system with slice scheduling and analyzed the scheduling rule with Pareto optimality from multi-objective Tabu search to prioritize applications. These studies may work nicely when encountering fixed scenarios but cannot reach the optimal slicing and offloading policy in highly-dynamic edge scenarios. Moreover, they typically require massive iterations, causing excessive computational overheads, and thus they cannot efficiently handle the large-scale problem of network slicing and computation offloading in complex multi-edge systems.

As an advanced intelligent algorithm, DRL automatically optimizes policies by interacting with unknown environments, which can be deemed as a promising solution. A few studies tried to customize DRL into the joint problem of network slicing and computation offloading. For instance, Liu *et al*. [19] designed refined network slices for V2X services and proposed a double DQN-based method for computation offloading and resource allocation. Chiang *et al*. [20] developed a DQN-based slicing framework to optimize slice scaling and computation offloading, aiming to enhance QoS and ESP profits. Shen *et al*. [26] designed a slicing-enabled offloading framework for vehicular networks by integrating service-oriented slicing with double DQN. However, when coping with complex offloading environments with huge decision spaces, the classic DRL cannot efficiently find the optimal solution and is often stuck in the local sub-optimum. Meanwhile, these studies commonly evaluate state-action pairs through using the maximized Q-value, resulting in the Q-value overestimation, and the policy might not perform steady convergence due to cumulative estimation errors and high variance. To the best of our knowledge, this is the first of its kind to propose a resource-efficient offloading framework that integrates dual-distillation DRL with self-attention-based prediction for addressing the joint problem of network slicing and computation offloading in dynamic multi-edge systems.

## 3 System Model and Problem Formulation

Fig. 1 illustrates the proposed computation offloading model towards traffic-cognitive network slicing in a multi-edge system. The system contains several edge regions, and the Base Station (BS) and edge servers in each edge region offer communication and computing resources for processing the offloaded tasks from the intelligent applications of users in the region, respectively. Users are randomly distributed within the communication coverage of each BS, and the ESP deploys services to edge servers in all regions to achieve service coverage for all users in the system. In each edge region, the ESP first sends slicing requests with demanded resources to the InP and makes payment. Next, the ESP deploys the offloading service on the edge server and manages the owned resources and system states. Finally, through paying fees, users can access the offloading service for processing their tasks.

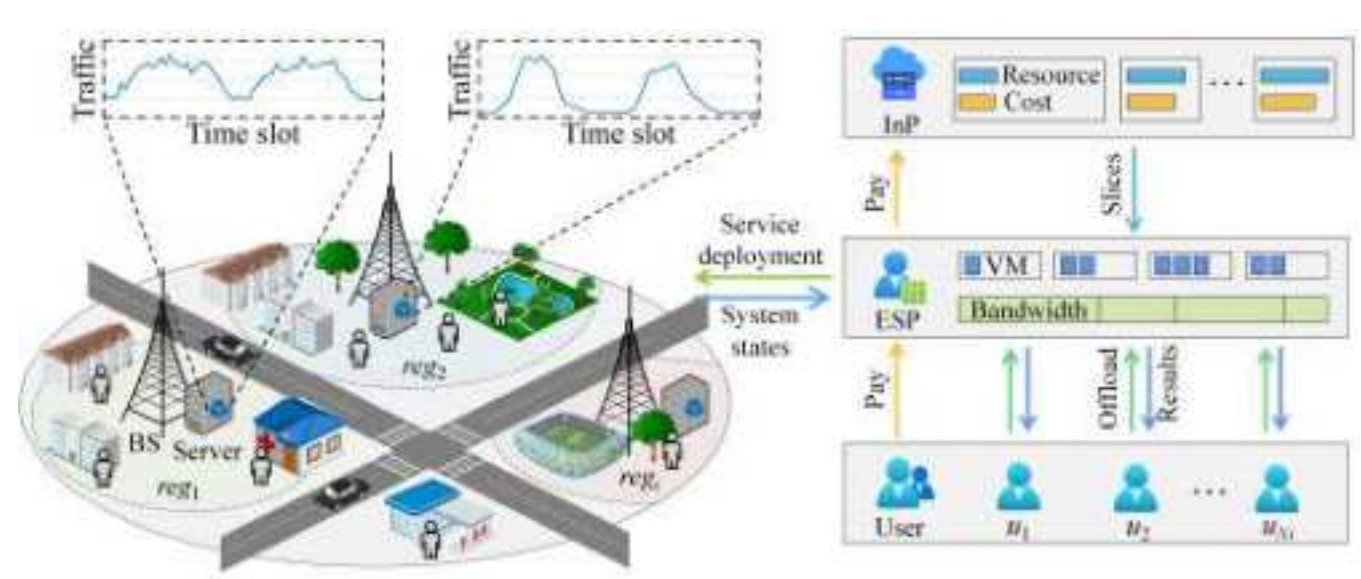

Fig. 1. The proposed computation offloading model towards traffic-cognitive network slicing in a multi-edge system.

Specifically, the edge regions in the system are defined as the set $Reg = \{reg_1, reg_2, ..., reg_i, ..., reg_{|Reg|}\}$, and the users within the coverage of the region $reg_i$ are denoted as the set $U_i = \{u_{i,1}, u_{i,2}, ..., u_{i,j}, ..., u_{i,N_i}\}$, where $N_i$ is the number of users in $reg_i$. The communication resources of BSs and the computing resources of edge servers are

provided by bandwidths and virtual machines (VMs), respectively. Due to user mobility and request randomness, the number of users sending offloading in different regions and time-slots experiences fluctuations, causing inconsistent spatio-temporal distributions of service demands.

To improve the efficiency of resource utilization and ESP profits, the ESP analyzes recent user traffic and their resource demands in each region at regular intervals, and then makes slice resource adjustments. To avoid QoS degradation caused by frequent slice adjustments, we adopt two scales of time-slots to cope with problems with different dimensions. Specifically, a long time-slot is denoted as $h$, where $h \in \{1, 2, ..., H\}$. At the start of $h$, the ESP evaluates the required slice resources in each region based on the information of user traffic and offloaded tasks collected in historical time-slots and sends slicing requests to the InP. Meanwhile, the long time-slot $h$ is divided into several short time-slots, and each is denoted as $t$, where $t \in \{1, 2, ..., T\}$. At the start of $t$, users upload their offloading requests to the ESP. The ESP assesses resource demands and user priorities and then makes proper decisions on computation offloading and resource allocation. At the end of $t$, the ESP collects the information of user traffic and task completion status in each region for subsequent slice adjustments.

### 3.1 Communication and Computation Models

The task of the user $u_{i,j}$ is clarified as a 4-tuple $< d_{i,j}, \eta_{i,j}, \rho_{i,j}, l_{i,j} >$, which are the data size, computing density, priority of $u_{i,j}$, and distance between $u_{i,j}$ and the BS, respectively. $\rho_{i,j}$ indicates the service level. More revenues can be obtained if the higher-priority tasks are completed.

When $u_{i,j}$ offloads its task to the edge server for execution, the input data should first be uploaded. The bandwidth allocated to $u_{i,j}$ is denoted as $b_{i,j}^{up}$, referring to Shannon's theorem [27], the rate of task uploading is

$$r_{i,j} = b_{i,j}^{up} \log_2(1 + \frac{p g_{i,j}}{\sigma^2}), \tag{1}$$

where $p$, $\sigma^2$ and $g_{i,j} = \beta_0 l_{i,j}^{-\theta}$ are the upload power, noise power, and channel power gain between $u_{i,j}$ and the BS, respectively.

Therefore, the task uploading time is defined as

$$T_{i,j}^{up} = \frac{d_{i,j}}{r_{i,j}}. \tag{2}$$

Once a task is uploaded to the BS, the ESP will schedule the task to a specific VM for execution. A VM may execute multiple tasks simultaneously, and thus it maintains a task waiting queue. When a task is scheduled to $VM_m$, the task queuing time is defined as

$$T_{i,j}^{que} = \sum_{k=1}^{|Q_m|} \frac{d_{i,k}\eta_{i,k}}{f^{edge}}, \tag{3}$$

where $Q_m$ represents the task waiting queue that already exists when a task arrives at $VM_m$, and $f^{edge}$ is the computing frequency of $VM_m$.

Therefore, the task execution time is defined as

$$T_{i,j}^{exe} = \frac{d_{i,j}\eta_{i,j}}{f^{edge}}. \tag{4}$$

After completing tasks, the results will be returned to users. Since the output data is much smaller than the input data, the result returning time is typically ignorable [28]. Through integrating the communication and computation models, the task completion time can be calculated by

$$T_{i,j}^{total} = T_{i,j}^{up} + T_{i,j}^{que} + T_{i,j}^{exe}. \tag{5}$$

### 3.2 Profit Model

When assessing ESP profits, both the revenues and costs of processing tasks should be considered. The ESP obtains revenues from users by offering services. If user tasks are completed within the maximum tolerable delay $T^{\max}$, the ESP will obtain the revenue $\Phi$. Otherwise, no revenue will be obtained. Hence, the revenue obtained from $u_{i,j}$ within $t$ can be described as

$$v_{i,j}^t = \begin{cases} \Phi, & T_{i,j}^{total} \leq T^{\max} \\ 0, & otherwise \end{cases}. \tag{6}$$

Further, the ESP will obtain different revenues if it completes tasks with diverse priorities. Therefore, the total revenues obtained in all regions within $h$ is defined as

$$R^h = \sum_{t=1}^{T} \sum_{i=1}^{|Reg|} \sum_{j=1}^{N_i} v_{i,j}^t \rho_{i,j}. \tag{7}$$

Meanwhile, the ESP will make payment for the rented resources in each region. The available bandwidths sold by InP in $reg_i$ are defined as $B_i = \{b_i^1, b_i^2, ...b_i^{|B_i|}\}$ with the costs of $\{\zeta_i^1, \zeta_i^2, ..., \zeta_i^{|B_i|}\}$. The available numbers of VMs sold by InP in $reg_i$ are defined as $V_i = \{v_i^1, v_i^2, ...v_i^{|V_i|}\}$ with the costs of $\{\xi_i^1, \xi_i^2, ..., \xi_i^{|V_i|}\}$. The bandwidths and VMs rented by ESP in the $reg_i$ are presented as

$$B_{esp}^i = \sum_{k=1}^{|B_i|} \alpha_i^k b_i^k, \; V_{esp}^i = \sum_{k=1}^{|V_i|} \beta_i^k v_i^k, \tag{8}$$

where $\alpha_i^k \in \{0, 1\}$ and $\beta_i^k \in \{0, 1\}$ are the renting decisions of the ESP for bandwidths and VMs in $reg_i$, respectively.

The total bandwidths and VMs rented by ESP in all regions at time-slot $h$ are

$$B_{esp}^h = \sum_{i=1}^{|Reg|} B_{esp}^i, V_{esp}^h = \sum_{i=1}^{|Reg|} V_{esp}^i. \tag{9}$$

Thus, the costs of renting resources within $h$ are

$$C^h = \sum_{i=1}^{|Reg|} \sum_{k=1}^{|B_i|} \alpha_i^k \zeta_i^k + \sum_{i=1}^{|Reg|} \sum_{k=1}^{|V_i|} \beta_i^k \xi_i^k. \tag{10}$$

### 3.3 Problem Formulation

Our objective is to maximize long-term ESP profits, and thus the optimization problem is formulated as

$$
\begin{aligned}
\text{P1:} \quad & \max_{\alpha,\beta,b,x} \sum_{h=1}^{H} (R^h - C^h) \qquad (11)\\
\text{s.t.} \quad & C1: \alpha_i^k, \beta_i^k \in \{0,1\}, \forall i, \forall k,\\
& C2: \sum_{k=1}^{|B_i|} \alpha_i^k = 1, \forall i,\\
& C3: \sum_{k=1}^{|V_i|} \beta_i^k = 1, \forall i,\\
& C4: \sum_{j=1}^{N_i} b_{i,j}^{up} \le B_{esp}^i, \forall i,\\
& C5: \sum_{j=1}^{N_i} f^{edge} \le V_{esp}^i, \forall i,
\end{aligned}
$$

where $\alpha$ and $\beta$ represent the bandwidth and VM renting decisions of ESP in all regions, respectively. $b$ and $x$ represent the bandwidth and VM allocation decisions in all regions, respectively. $C1$ indicates that ESP can only decide whether to fully rent a type of resource. $C2$ and $C3$ indicate that the ESP can only rent a type of bandwidth and VM. $C4$ and $C5$ indicate that the allocated bandwidths and VMs cannot exceed the resources rented by ESP.

**Theorem 1.** *P1 is an NP-hard problem.*

**Proof.** We aim to prove a special case of $P1$ is equivalent to the Maximum Budget Coverage Problem (BMCP) that is NP-hard [29]. In BMCP, there is a set $E = \{e_1, e_1, ..., e_n\}$, where each element owns specific value and cost. The objective of the BMCP is to find the subset $E' \subseteq E$ that can maximize total values while meeting the cost constraint. In the edge region $reg_i$, when $\alpha$, $\beta$, and $b$ in $P1$ are fixed, the offloading requests can be deemed as the elements of $E$, where the allocated VMs and the revenues from completing tasks are mapped to costs and values, respectively. Thus, this special case of $P1$ is defined as

$$
\begin{aligned}
& \max \sum_{j=1}^{N_i} v_{i,j}^t \rho_{i,j}\\
& \text{s.t.} \sum_{j=1}^{N_i} f^{edge} \le V_{esp}^i.
\end{aligned} \qquad (12)
$$

It can be found that this special case is equivalent to the NP-hard BMCP. By extending the above special case to a multi-edge system, we can derive that $P1$ is NP-hard.

To enhance QoS and ESP profits, it is essential to properly adjust slice resources during the offloading process. The increase or decrease of user traffic and slice resources might lead to significant changes in problem space. Moreover, the problems of network slicing and computation offloading belong to different time scales, raising the problem-solving difficulty. To relieve this issue, we decouple $P1$ into two sub-problems and formulate them as follows.

- $P1.1$: This sub-problem is to minimize ESP resource costs in long time-slots by making slice adjustments while meeting user demands, which is defined as

$$
\begin{aligned}
\text{P1.1:} \quad & \min_{\alpha,\beta} \sum_{h=1}^{H} C^h \qquad (13)\\
\text{s.t.} \quad & C1 - C3,\\
& C6: T_{i,j}^{total} \le T^{max}, \forall i, \forall j.
\end{aligned}
$$

- $P1.2$: This sub-problem is to maximize the ESP revenues in short time-slots by conducting computation offloading and resource allocation, which is defined as

$$
\begin{aligned}
\text{P1.2:} \quad & \max_{b,x} \sum_{t=1}^{T} \sum_{i=1}^{|Reg|} \sum_{j=1}^{N_i} v_{i,j}^t \rho_{i,j} \qquad (14)\\
\text{s.t.} \quad & C4 - C5.
\end{aligned}
$$

## 4 The Proposed *SliceOff*

In this section, we first present an overview of the *SliceOff*. Next, we describe the two core components of the *SliceOff* in detail and provide rigorous theoretical proofs for the effectiveness. Finally, we analyze the complexity of the *SliceOff*.

### 4.1 Overview of the *SliceOff*

The proposed *SliceOff* aims to improve ESP profits by jointly solving the sub-problems of network slicing ($P1.1$) and computation offloading ($P1.2$). For $P1.1$, linear programming and random rounding are adopted to obtain the optimal slicing scheme. To enhance the cognition ability for fluctuating user traffic, a self-attention-based traffic prediction model is devised to support adaptive slice adjustments. For $P1.2$, we develop an improved DRL method, where a dual-distillation mechanism is designed to explore the optimal policy from different environmental perspectives to enhance the learning efficiency and adaptability of DRL agents in huge problem spaces.

Fig. 2 illustrates the overview of the *SliceOff*, where the main workflow is outlined in Algorithm 1. For each long time-slot, Algorithm 2 is first called to obtain the renting decisions of bandwidth and VM of each region (i.e., $\alpha_i^k$ and $\beta_i^k$), and then perform slice adjustments and calculate resource costs based on total renting resource (i.e., $B_{esp}^h$ and $V_{esp}^h$) (Lines 2∼3). At the start of each short time-slot, offloading requests are first sent to the ESP and the task information is uploaded to the BS (Line 5). Next, Algorithm 3 is called to generate bandwidth allocation and offloading decisions (i.e., $b^t$ and $x^t$ ) of each region. Then, the offloaded tasks are executed on edge servers according to $b^t$ and $x^t$ and the results are returned after task completion (Lines 6∼7). At the end of each short time-slot, the ESP calculates the revenues based on task completion time and user priority and collects the state information of each region for subsequent slice resource adjustments (Lines $8 \sim 9$). Finally, it goes to the next long time-slot.

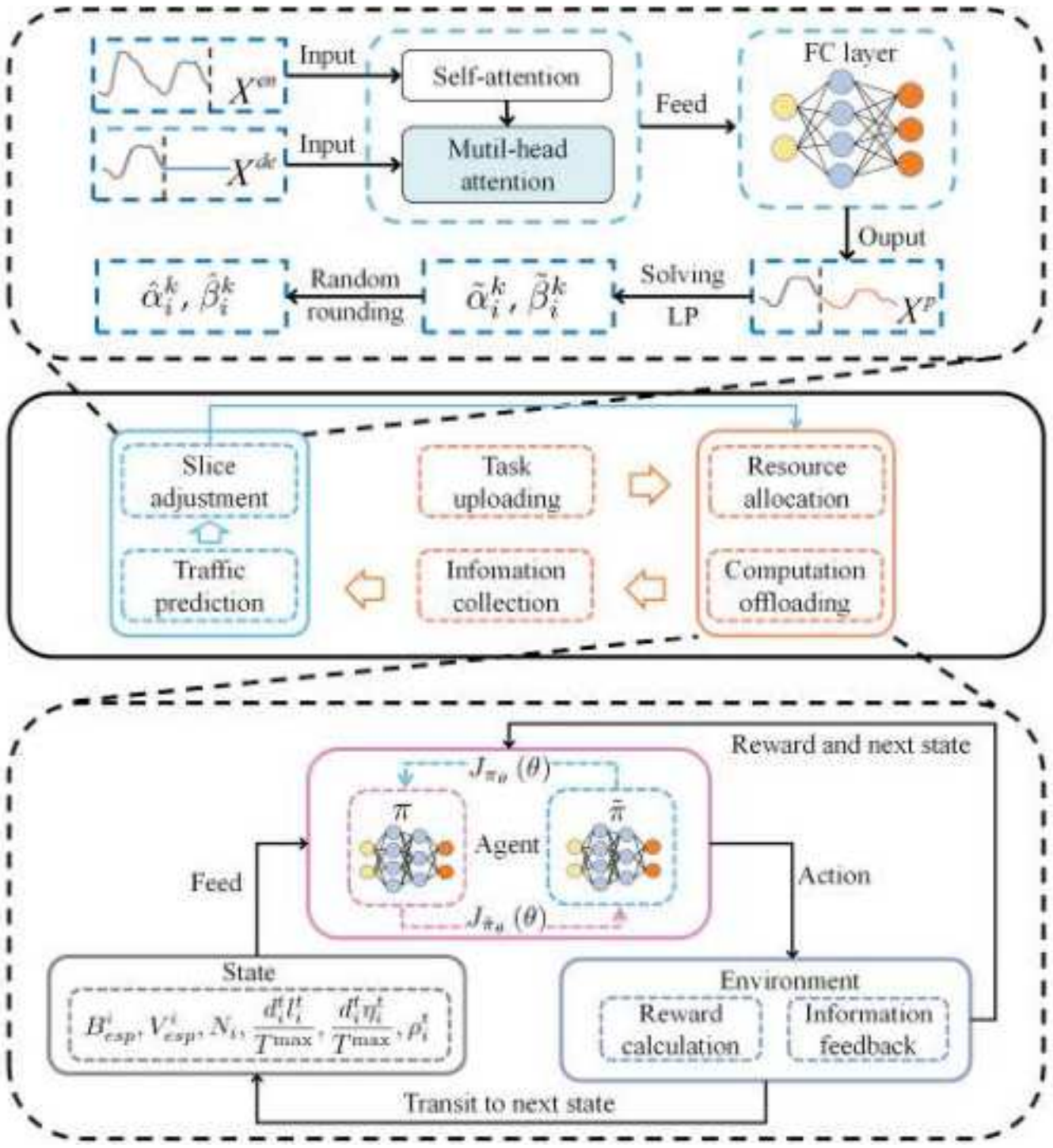

Fig. 2. Overview of the proposed *SliceOff*.

**Algorithm 1:** The proposed *SliceOff*

**Input:** $B$, $V$

**Output:** Slicing and offloading policies

**for** $h$ = 1, 2, ..., $H$ **do**
- Call **Algorithm 2** to obtain $\alpha_i^k$ and $\beta_i^k$;
- Perform slice adjustments and calculate resource costs based on $B_{esp}^h$ and $F_{esp}^h$;
- **for** $t$ = 1, 2, ..., $T$ **do**
  - Send offloading requests to the ESP;
  - Call **Algorithm 3** to generate $b^t$ and $x^t$;
  - Execute tasks and return results;
  - Calculate ESP revenues;
  - Collect the state information of each region;

## 4.2 Prediction-assisted Slice Adjustment

By evaluating user traffic and offloading demands in various regions, slices can be adjusted in advance to better meet user demands and improve resource efficiency. In multi-edge environments, the traffic exhibits long-term changes (impacted by application popularity) and short-term fluctuations (impacted by user mobility), and meanwhile, different edge regions may have various traffic patterns. These factors pose significant challenges to traffic prediction. Compared to classic Recurrent Neural Networks (RNNs) and Convolutional Neural Networks (CNNs), the emerging Transformer [30] introduces a self-attention mechanism to capture the temporal dependencies in the sequence, avoiding the problem of gradient explosion or vanishing. Considering its excellent ability, we design a self-attention-based method to predict user traffic in different regions. Moreover, service demands commonly depend on the services provided by the ESP, which can be estimated by analyzing historical task attributes and system loads. Specifically, we adopt a linear programming (LP) solver to obtain optimal slicing resources based on traffic prediction and service demands analysis. Considering that $P1.1$ is constrained by $C1$ (i.e., $\alpha_i^k, \beta_i^k \in \{0,1\}$) and the solver outputs real solutions, we utilize random rounding to round real solutions to feasible ones for making slice-adjustment decisions. The key steps of the proposed adaptive slice adjustment method are given in Algorithm 2.

**Algorithm 2:** Prediction-assisted slice adjustment

**Input:** Historical system states, $B$, $V$

**Output:** $\hat{\alpha}_i^k$, $\hat{\beta}_i^k$

1 Construct input vectors using historical user traffic: $X^{en} \leftarrow X^{his}$, $X^{de} \leftarrow Concat\left(X^{cur}, X^0\right)$;
2 Output of encoder: $H_1 = Encoder(X^{en})$;
3 Output of decoder: $H_2 = Decoder(H_1, X^{de})$;
4 Predict future traffic: $X^p = MLP(H_2)$;
5 Convert $C6$ of $P1.1$ into Eq. (17);
6 Construct LP solver by relaxing $C1$ to $\alpha_i^k$ and $\beta_i^k \in [0, 1]$;
7 Obtain the optimal $\tilde{\alpha}_i^k$ and $\tilde{\beta}_i^k$;
8 **for** $i$ = 1, 2, ..., $|Reg|$ **do**
9     **for** $k$ = 1, 2, ..., $|B_i|$ **do**
10         Round $\hat{\alpha}_i^k = 1$ or $\hat{\alpha}_i^k = 0$ according to $\tilde{\alpha}_i^k$;
11     **for** $k$ = 1, 2, ..., $|V_i|$ **do**
12         Round $\hat{\beta}_i^k = 1$ or $\hat{\beta}_i^k = 0$ according to $\tilde{\beta}_i^k$;

*Step 1: Predict future traffic.* The user traffic in historical system states is first used to construct the input vectors of the encoder and decoder (Line 1). $X^{his}$ indicates the historical traffic of all regions. $X^{cur}$ indicates the traffic collected in the current prediction window that contains the traffic from previous long time-slots. To avoid the high complexity caused by calculating the attention weights of all historical time-slots, we adopt a probsparse self-attention [31] in the encoder and design a self-attention distillation between layers to reduce network overheads (Line 2). Specifically, the feature extraction from $j$-th to $(j+1)$-th layers is defined as

$$X_{j+1}^{en} = \text{MaxPool}(\text{ELU}(\text{Conv1d}([X_j^{en}]_{attention}))), \quad (15)$$

$$[\cdot]_{attention} = \text{Softmax}(\frac{\bar{Q}K^{\mathrm{T}}}{\sqrt{d}})V, \quad (16)$$

where $[\cdot]_{attention}$ indicates the probsparse self-attention, $d$ is the dimension of $X_j^{en}$, Conv1d indicates the one-dimensional convolution, ELU is the activation function, and MaxPool indicates the maximum pooling. $\bar{Q}$ is a sparse matrix that contains the Top-u queries, making the probsparse self-attention calculating only $O(\log L)$ dot-product for each query-key lookup, where $L$ is the length of $X^{cur}$.

Next, the output of the encoder and $X^{cur}$ are fed into the decoder, which consists of a multi-head probsparse self-attention and a masked multi-head attention (Line 3). Finally, the output of the decoder is fed into the Fully Connected (FC) layer, and thus the future traffic in all regions (i.e., $X^p$) can be predicted (Line 4).

*Step 2: Obtain the relaxed optimal slicing resource solution.* We convert $C6$ of $P1.1$ into the resource demands for constructing a convex objective function (Line 5). Considering that $T^{max}$ depends on the bandwidth and computing resources allocated to users, we introduce a delay ratio $\omega$ to split $T^{max}$. Specifically, the converted $C6$ is defined as

$$\frac{B_{esp}^{i}}{N_i}H(1+SNR) \geq \frac{\bar{d}}{\omega T^{\max}}, \frac{V_{esp}^{i}}{N_i} \geq \frac{\bar{d}_i \bar{\eta}_i}{(1-\omega)T^{\max}}, \quad (17)$$

where $\bar{d}_i$ and $\bar{\eta}_i$ indicate the expected data size and computing density of tasks in $reg_i$, which can be attained by analyzing historical task information. $\omega$ is the ratio between communication delay and total delay, which depends on the provided offloading services and the distribution of task attributes in $reg_i$.

Based on the above conversion and traffic prediction, we construct an LP solver to minimize resource costs while satisfying the delay constraint. Since $\alpha_i^k$ and $\beta_i^k$ must be integer solutions, we first relax $C1$ to $\alpha_i^k$ and $\beta_i^k \in [0,1]$. Next, the predicted traffic, task demands, and resource costs are input into the solver to obtain the optimal renting solutions of bandwidths $\tilde{\alpha}_i^k$ and VMs $\tilde{\beta}_i^k$ (Lines 6∼7).

*Step 3: Randomly round for the feasible slicing resource solution.* To keep $C1$, the optimal solution needs to be rounded. We regard the optimal solution as a probability and randomly round it to obtain the integer solutions $\hat{\alpha}_i^k = 1$ or $\hat{\alpha}_i^k = 0$ (Lines 9∼10). To satisfy $C2$, there must be one $\hat{\alpha}_i^k = 1$ for $reg_i$ and the rest are 0. Similarly, the feasible solutions of $\hat{\beta}_i^k$ are obtained according to $\tilde{\beta}_i^k$ (Lines 11∼12).

**Lemma 1.** (Chernoff Bound [32]) There are $n$ independent variables $\{x_1, x_2, ..., x_n\}$, where $\forall x_i \in [0,1]$. Let $\mu = \mathbb{E}[\sum_{i=1}^{n} x_i]$, $\mathbf{Pr}[\sum_{i=1}^{n} x_i \geq (1+\epsilon)\mu] \leq e^{\frac{-\epsilon^2\mu}{2+\epsilon}}$ and $\mathbf{Pr}[\sum_{i=1}^{n} x_i \leq (1-\epsilon)\mu] \leq e^{\frac{-\epsilon^2\mu}{2}}$, where $\varepsilon$ is an arbitrarily positive value.

**Theorem 2.** *Algorithm 2 approximates the optimal resource cost derived by solving the LP with a high probability.*

**Proof.** With $\omega$, the optimal $B_{esp}^h$ and $V_{esp}^h$ can be achieved separately, and thus we regard $B_{esp}^h$ as an example to prove. According to Eq. (8), the lowest bandwidth cost obtained by the LP is $\tilde{B}_{esp}^h = \sum_{i=1}^{|Reg|} \tilde{B}_{esp}^i = \sum_{i=1}^{|Reg|} \sum_{k=1}^{|B_i|} \tilde{\alpha}_i^k b_i^k$. After rounding, we can obtain the feasible solution $\hat{\alpha}_i^k$. Therefore, the rounded bandwidth cost is calculated by

$$\hat{B}_{esp}^i = \begin{cases} b_i^k, & with\ \tilde{\alpha}_i^k \\ 0, & otherwise \end{cases}. \quad (18)$$

It is noted that $E[\sum_{i=1}^{|Reg|} \hat{B}_{esp}^i] = \sum_{i=1}^{|Reg|} \sum_{k=1}^{|B_i|} \tilde{\alpha}_i^k b_i^k = \sum_{i=1}^{|Reg|} \tilde{B}_{esp}^i$. Next, we define $B_{\max}$ as the maximum bandwidth in all regions, and thus $E[\frac{\sum_{i=1}^{|Reg|} \hat{B}_{esp}^i}{|Reg|B_{\max}}] \in [0,1]$. For $reg_i$, $\hat{B}_{esp}^i$ is independent of other regions, which satisfies the constraints in Lemma 1. Thus, we can derive

$$Pr[\frac{\sum_{i=1}^{|Reg|} \hat{B}_{esp}^i}{|Reg|B_{\max}} \geq (1+\epsilon)E[\frac{\sum_{i=1}^{|Reg|} \hat{B}_{esp}^i}{|Reg|B_{\max}}]] \leq e^{\frac{-\epsilon^2}{2+\epsilon} E[\frac{\sum_{i=1}^{|Reg|} \hat{B}_{esp}^i}{|Reg|B_{\max}}]}$$
$$\Leftrightarrow Pr[\sum_{i=1}^{|Reg|} \hat{B}_{esp}^i \geq (1+\epsilon)\sum_{i=1}^{|Reg|} \tilde{B}_{esp}^i] \leq e^{\frac{-\epsilon^2}{2+\epsilon} \frac{\sum_{i=1}^{|Reg|} \tilde{B}_{esp}^i}{|Reg|B_{\max}}}. \quad (19)$$

Since $\varepsilon$ is an arbitrarily positive value, there is only a very low probability for Algorithm 2 to exceed the bandwidth cost achieved by the LP. Therefore, $B_{esp}^h = \sum_{i=1}^{|Reg|} \hat{B}_{esp}^i$ has a high probability of approximating the optimal solution $\sum_{i=1}^{|Reg|} \tilde{B}_{esp}^i$ derived by solving the LP.

### 4.3 Improved DRL with Dual Distillation for Computation Offloading and Resource Allocation

When dealing with the complex problem of computation offloading and resource allocation with variable available resources, existing DRL-based methods reveal the performance bottlenecks caused by the Q-value overestimation and low exploration efficiency. To address these issues, we propose an improved DRL method with dual distillation, whose main steps are described in Algorithm 3. Specifically, we introduce twin critics' networks with a delay mechanism to solve the Q-value overestimation and reduce variance. Meanwhile, inspired by the dual distillation [33], we adopt two DRL agents that conduct explorations from different environmental perspectives and distill knowledge from each other to improve the learning efficiency in huge decision spaces. We consider the problem model of $P1.2$ as the environment, and each DRL agent interacts with the environment to optimize its policy while distilling knowledge from the peer agent's policy. The state space, action space, and reward function are defined as follows.

**Algorithm 3:** Improved DRL with dual distillation for computation offloading and resource allocation

```
   Input: B^i_esp, V^i_esp
   Output: b^t, x^t
1  Initialize: current and peer DRL agents
2  for epoch = 1, 2, ..., E do
3      Initialize state: s_0 = env.reset();
4      for t = 1, 2, ..., T do
5          Explore action of computation offloading and
             resource allocation: a_t = π(s_t|θ^π) + N_t;
6          Feedback r_t and s_{t+1} after executing a_t:
             r_t, s_{t+1} = env.step(a_t);
7          Store state-transition samples in replay
             buffer: RB.push(s_t, a_t, r_t, s_{t+1});
8          Randomly select K samples:
             K * (s_t, a_t, r_t, s_{t+1})= RB.Sample(K);
9          Obtain ã_{t+1} at s_{t+1} according to Eq. (23);
10         Calculate target Q-value y_t by using r_t
             according to Eq. (24);
11         Update Q_1 and Q_2:
             φ^{Q_i} ← min(y_t − Q_i(s_t, a_t));
12         if t mod 2 = 0 then
13             Update actor's network by Eq. (25);
14             Update actor's network with π̃ by Eq. (26);
15             Update target networks via soft update;
16     Update peer agent π̃;
```

- **State space.** It comprises the available bandwidth, VMs, user traffic, and task attributes in the current time-slot. To better capture demand features, we transfer the data size and computing density of tasks into the demands of uploading rate and computing frequency. Thus, the system state at $t$ in $reg_i$ is defined as

$$s_t = \{B_{esp}^i, V_{esp}^i, N_i, \frac{d_i^t l_i^t}{T^{\max}}, \frac{d_i^t \eta_i^t}{T^{\max}}, \rho_i^t\}, \tag{20}$$

where $d_i^t$, $l_i^t$, $\eta_i^t$, and $\rho_i^t$ are vectors. For example, $d_i^t = \{d_{i,1}^t, d_{i,2}^t, \ldots, d_{i,N_i}^t\}$.

- **Action space.** The DRL agent should simultaneously determine the bandwidths and VMs to be allocated, and thus the action at $t$ is defined as

$$a_t = \{b_i^t, x_i^t\}, \tag{21}$$

where $b_i^t = \{b_{i,1}^t, b_{i,2}^t, \ldots, b_{i,N_i}^t\}$ indicates the proportion of bandwidths, and the bandwidth allocated to $u_{i,j}$ is $b_{i,j}^t B_{esp}^i$. $x_i^t = \{x_{i,1}^t, x_{i,2}^t, \ldots, x_{i,N_i}^t\}$ indicates the VM index, and the VM index allocated to $u_{i,j}$ is $\left\lfloor x_{i,j}^t V_{esp}^i \right\rfloor$.

- **Reward function.** The objective of solving $P1.2$ is to maximize cumulative ESP revenues, and thus the reward function is defined as

$$r_t = \sum_{j=1}^{N_i} \upsilon_{i,j}^t \rho_{i,j}. \tag{22}$$

In Algorithm 3, we first initialize the current and peer DRL agents (Line 1), where each agent consists of online networks (two critics' networks $Q_1$ and $Q_2$ and actor's network $\pi$) and target networks (${Q_1}'$, ${Q_2}'$, and $\pi'$). Different from classic DRL that directly employs the maximized Q-value, the proposed method adopts two independent critics' networks to fit the Q-value function. This design alleviates Q-value overestimation and avoids getting stuck in the sub-optimum due to undeserved cumulative errors. For each training epoch, the environment is first initialized for the current DRL agent (Line 3). At each short time-slot $t$, the state $s_t$ is fed into the actor's network $\pi$, and then the DRL agent explores the action $a_t$ at the current state according to $\pi$ and exploration noise (Line 5). After executing computation offloading and resource allocation, the environment feedbacks the immediate reward $r_t$ and next state $s_{t+1}$ (Line 6). Next, the state-transition samples are stored in a replay buffer (Line 7), and then $K$ samples are randomly selected to update network parameters (Line 8). When updating the critic's network, $\tilde{a}_{t+1}$ at $s_{t+1}$ is first obtained by the target actor's network (Line 9). This process is defined as

$$\tilde{a}_{t+1} = \pi'(s_{t+1}|\theta^{\pi'}) + \varepsilon, \ \varepsilon \sim N(0, \sigma), \tag{23}$$

where the network noise $\varepsilon$ is utilized as regularization that makes similar actions hold equivalent rewards.

Next, the target Q-value is calculated by using $r_t$ and comparing two critics' networks (Line 10), which is described as

$$y_t = r(s_t, a_t) + \gamma \cdot \min_{i=1,2} ({Q'}_{\hat{\phi}_{i,j}}(s_{t+1}, \tilde{a}_{t+1})). \tag{24}$$

Then, the two critic's networks are updated (Line 11). To decrease the updating frequency of the policies with low quality, we adopt a delay mechanism to update the actor's network and target networks. If $t \bmod 2 = 0$, the actor's network will be updated by gradient ascent (Line 13). This process is defined as

$$\nabla_\theta J(\theta) = \frac{1}{K} \sum_{j=1}^{K} \nabla_a Q_1(s,a)|_{a=\pi(s)} \nabla_\theta \pi_\theta(s). \tag{25}$$

Next, the policy of the current agent is updated with the peer agent's policy $\tilde{\pi}$. However, it is hard to determine which of the two agents performs better at different states. To address this issue, we soften the loss function and introduce a weighted loss function to update the policy (Line 14). Specifically, the distillation objective function is defined as

$$J_{\tilde{\pi}}(\theta) = \mathbb{E}_{s\sim\tilde{\pi}} \left[ \| \pi(s) - \tilde{\pi}(s) \|_2^2 \exp(\alpha \xi^{\tilde{\pi}}(s)) \right], \tag{26}$$

where $\exp(\alpha \xi^{\tilde{\pi}}(s))$ is the confidence score, and $\alpha$ controls the confidence level that depends on the accuracy of value-function estimation. $\xi^{\tilde{\pi}}(s)$ indicates the advantage value of $\tilde{\pi}$ compared to $\pi$ at $s$, which is defined as

$$\xi^{\tilde{\pi}}(s) = V^{\tilde{\pi}}(s) - V^{\pi}(s), \tag{27}$$

where the larger $\xi^{\tilde{\pi}}(s)$ indicates the loss function is more conducive for the current policy to learn from the peer one. Otherwise, it is more inclined to remain unchanged.

Next, the target networks are updated via soft update (Line 15), and thus the critic's network is updated more frequently than the actor's and target networks. Unlike directly updating all networks, this manner reduces cumulative errors and improves training stability. Finally, the algorithm is continuously running to update the peer agent's policy (Line 16). During this process, each agent is updated according to its reward function and distillation loss function, achieving its own update while learning useful knowledge from the peer agent to enhance itself.

**Theorem 3.** *Dual distillation between the current and peer agents supports achieving a more optimized hybrid policy.*

**Proof.** A hybrid policy consists of the current and peer agents' policies that are chosen according to the relative advantage value, and it is defined as

$$\pi^*(s) = \begin{cases} \pi(s), & \xi^{\tilde{\pi}}(s) > 0 \\ \tilde{\pi}(s), & otherwise \end{cases}. \tag{28}$$

Therefore, it can be guaranteed that $\pi^*$ is a more optimized policy than $\pi$ and $\tilde{\pi}$ (i.e., $\forall s$, $V^{\pi^*}(s) \geq V^{\pi}(s)$ and $V^{\pi^*}(s) \geq V^{\tilde{\pi}}(s)$). Next, we consider a simple form of the objective function, which is defined as

$$J'_{\tilde{\pi}}(\theta) = \mathbb{E}_{s\sim\tilde{\pi}}[||\pi(s) - \tilde{\pi}(s)||^2 \mathbb{1}(\xi^{\tilde{\pi}}(s) > 0)], \tag{29}$$

where $\mathbb{1}$ is the indicator function. If $\xi^{\tilde{\pi}}(s) > 0$, the value of $\mathbb{1}(\cdot)$ is 1. Otherwise, this value is 0.

Since $\rho_\pi$ and $\rho_{\tilde{\pi}}$ is similar, the difference between $\rho_{\tilde{\pi}}$ and $\rho_{\pi^*}$ is negligible. Thus, the proposed dual-distillation process can be described as

$$\begin{aligned} &\mathbb{E}_{s\sim\tilde{\pi}}[||\pi(s) - \tilde{\pi}(s)||^2 \mathbb{1}(\xi^{\tilde{\pi}}(s) > 0)] \\ &= \sum_{s\sim\rho_{\tilde{\pi}};\xi^{\tilde{\pi}}(s)>0} ||D(\pi(s) - \tilde{\pi}(s))||^2 + \sum_{s\sim\rho_{\tilde{\pi}};\xi^{\tilde{\pi}}(s)\leq 0} ||D(\pi(s) - \pi(s))||^2 \\ &= \sum_{s\sim\rho_{\pi^*};\xi^{\tilde{\pi}}(s)>0} ||D(\pi(s) - \tilde{\pi}(s))||^2 + \sum_{s\sim\rho_{\pi^*};\xi^{\tilde{\pi}}(s)\leq 0} ||D(\pi(s) - \pi(s))||^2 \\ &= \sum_{s\sim\rho_{\pi^*}} ||\pi(s) - \pi^*(s)||^2 \\ &= \mathbb{E}_{s\sim\pi^*}[||\pi(s) - \pi^*(s)||^2]. \end{aligned} \tag{30}$$

Based on the above proof, we can derive that the knowledge of $\tilde{\pi}$ will be transferred to $\pi$ if the advantage value of $\tilde{\pi}$ is positive. Otherwise, the update of the current policy will ignore the dual-distillation process. Therefore, the dual-distillation enables a more optimized hybrid policy. Further, to reduce the errors of value-function estimation caused by neural networks, we replace $\mathbb{1}(\xi^{\tilde{\pi}}(s))$ with $\exp(\xi^{\tilde{\pi}}(s))$ and use $\alpha$ to control the confidence level for obtaining the objective function.

### 4.4 Complexity Analysis

The proposed *SliceOff* consists of two main phases including offline training and online decision-making.

- **Offline training.** When there are $M$ training iterations in Algorithm 2, each iteration contains $H$ long time-slots, and the length of prediction window is $L_p$, the complexity of training Algorithm 2 is $O(MHL_p)$. When there are $E$ training epochs in Algorithm 3, each epoch contains $H$ long time-slots, each long time-slot contains $T$ short time-slots, and each short time-slot contains up to $N_{max}$ users, the complexity of training Algorithm 3 is $O(EHTN_{max})$.
- **Online decision-making.** For each long time-slot, the complexity of traffic prediction is $O(L_p)$, the complexity of slice adjustments is $O(\sum_{i=1}^{|Reg|}(|B_i|+|V_i|))$, the complexity of offloading decisions is $O(N_{max})$, and thus the complexity of the online decision-making is $O((L_p+\sum_{i=1}^{|Reg|}(|B_i|+|V_i|))/T+N_{max})$.

The number of regions served by ESP and the types of resources provided by InP are usually small, thus the *SliceOff* owns low complexity and linearly increases with user traffic. When $T$ is smaller, there will be more accurate traffic prediction and slice adjustments, but meanwhile, it will lead to higher complexity and more frequent service interruptions.

## 5 Performance Evaluation

In this section, we evaluate the proposed *SliceOff* by conducting extensive simulation and testbed experiments.

### 5.1 Experiment Setup

**Experimental environment and datasets.** Based on a workstation equipped with an 8-core Intel(R) Xeon(R) Silver 4208 CPU@3.2GHz, two NVIDIA GeForce RTX 3090 GPUs, and 32GB RAM, we construct the proposed system and implement the *SliceOff* by PyTorch. The system contains three edge regions (i.e., R1, R2, and R3) and 24 long time-slots, where each long time-slot contains 6 short time-slots. The real-world datasets of Milan cellular traffic [34] are adopted to simulate the offloading requests from users, which comprise three types of services (i.e., message, call, and Internet). The traffic during 2 months was recorded with the sampling frequency of 10 min. We select the traffic records of Internet services in the three regions (i.e., ID=4259, 4456, and 5060) as the number of user requests in the above three edge regions in a short time-slot.

**Parameter settings.** In each edge region, the user traffic fluctuates in [2, 10], $f^{edge}$ = 2.0 GHz, $p$ = 100 mW, $\beta_0$ = -60 dB, $\theta$ = 2, and $\sigma^2$ = -110 dBm [35]. For task attributes, $\rho_{i,j} \in \{1,2,3\}$, $l_{i,j} \in$ [1, 2000] meters, $T^{\max}$ = 1.0 s, and $\Phi$ = 1.0 $. The other parameters of different edge regions are listed in Table 1. For the parameters of the *SliceOff*, $\omega$ = 0.3, $\alpha$ = 1.0, and the learning rate is 0.001 [36].

TABLE 1
Parameter Settings of Different Edge Regions

| Region | R1 | R2 | R3 |
|---|---|---|---|
| $d_{i,j}$(KB) | [100,300] | [300,500] | [500,700] [37] |
| $\eta_{i,j}$ (cycles/bit) | [400,600] | [100,200] | [50,100] [37] |
| $B_i$ | [1,2,...,20] | [1,2,...,25] | [1,2,...,30] |
| $\zeta$ ($/$h$) | [3,6,...,60] | [2,4,...,50] | [1,2,...,30] [38] |
| $V_i$ | [1,2,...,16] | [1,2,...,12] | [1,2,...,8] |
| $\xi$ ($/$h$) | [2,4,...,32] | [4,8,...,48] | [6,12,...,48] [38] |

**Performance metrics.** Besides ESP profits, we consider the following metrics to further evaluate the *SliceOff*.

- Resource Utilization (RU): It is the rate between the actually-used resources for executing tasks and the resources rented by the ESP.
- Deadline Violate Rate (DVR): It is the rate between the number of tasks whose execution time surpasses the maximum tolerable delay and the total number of tasks.

**Comparison methods.** To verify the superiority of the *SliceOff*, we compare it with the following state-of-the-art methods for traffic prediction, slicing, and offloading.

- *GLTP* [39]: A global-local traffic prediction method was designed to make network slicing, which adopted temporal convolutional networks to learn regional traffic.
- *PredRNN* [40]: A predictive RNN-based traffic prediction method was designed to make network slicing, which adopted a gate mechanism to capture the dependencies in traffic sequences.
- *HRF-TD3* [41]: A twin-delayed deep deterministic policy gradient algorithm was designed to make offloading and resource allocation, which adopted twin critics' networks to evaluate Q-values.
- *PD-DDPG* [42]: A deep deterministic policy gradient algorithm was designed to make offloading and resource allocation, which adopted a critic's network.
- *Off*: There is no dynamic network slicing and it adopts the same offloading method as the *SliceOff*.

### 5.2 Experiment Results and Analysis

**Traffic distribution and prediction.** We analyze the diverse traffic changes in different regions and evaluate the prediction performance of the *SliceOff*. Specifically, Fig. 3 (a) shows the traffic distributions in different regions and the prediction accuracy of the *SliceOff*. In each region, the traffic fluctuates over time, and there are significant differences in the traffic range and fluctuation magnitude among different regions. When facing this complicated problem, the proposed traffic prediction model in *SliceOff* can well track traffic changes in each region and accurately predict future trends, which provides effective support for further slicing resource adjustments. Fig. 3 (b) illustrates how the *SliceOff* makes slice adjustments with traffic prediction in a region. When the traffic decreases (from around the time-slot 20), the *SliceOff* reduces the slice resources rented by the ESP to save system costs. When the traffic increases (from around the time-slot 100), the *SliceOff* accurately captures this trend and adds the slice resources to better meet user demands. Through such adaptive slice adjustments, the *SliceOff* can better meet user demands while avoiding unnecessary resource waste.

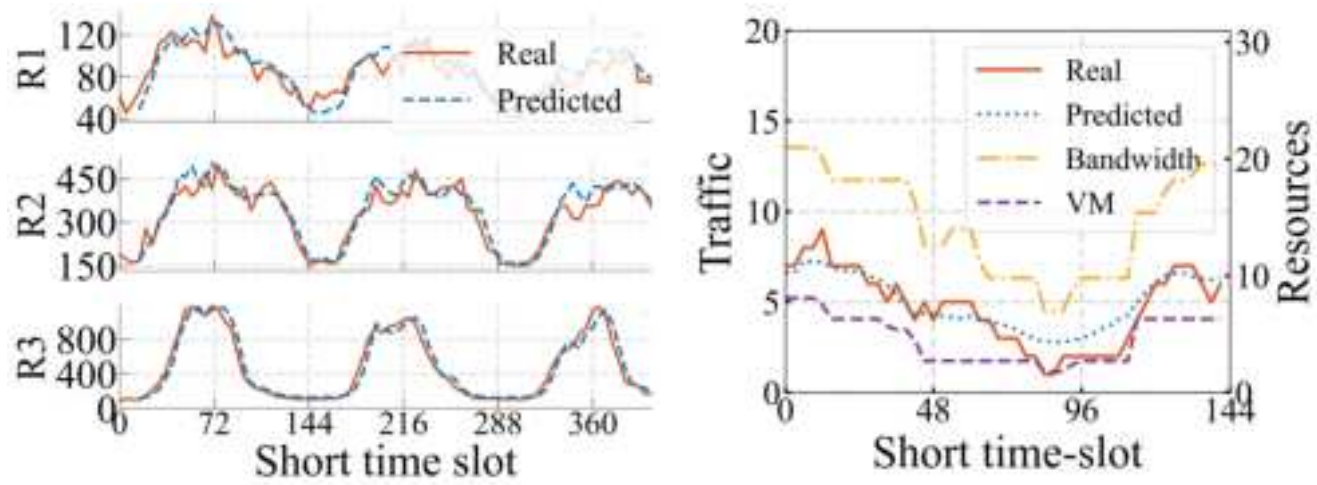

(a) Diverse traffic changes in different regions. (b) Adaptive slice adjustments with accurate traffic prediction.

Fig. 3. Diverse traffic changes in different regions and adaptive slice adjustments with accurate traffic prediction.

**Convergence of different methods.** We compare the convergence of the *SliceOff* with other methods. As depicted in Fig. 4, the *HRF-TD3* achieves higher rewards and more steady convergence than the *PD-DDPG*. This is because the *HRF-TD3* adopts two independent critics' networks, curing the problem of Q-value overestimation in the *PD-DDPG*. Meanwhile, the *HRF-TD3* employs a delay mechanism to avoid frequent network updates, lowering cumulative errors and thus enhancing training stability. Compared to the *HRF-TD3* and *PD-DDPG*, the *Off* converges to higher rewards. This is because the *Off* utilizes two DRL agents to explore the rewards in the environments and transfer knowledge to each other, which increases the training efficiency of the agents. By introducing dynamic slice adjustments, the *SliceOff*, *GLTP*, and *PredRNN* significantly enhance the rewards compared with the other three methods. This is because accurate traffic prediction can assist in performing dynamic slice resource adjustments to improve resource availability when the traffic is changeable, thereby increasing the rewards. The above results demonstrate the superiority of the proposed offloading framework with prediction-assisted slice adjustments.

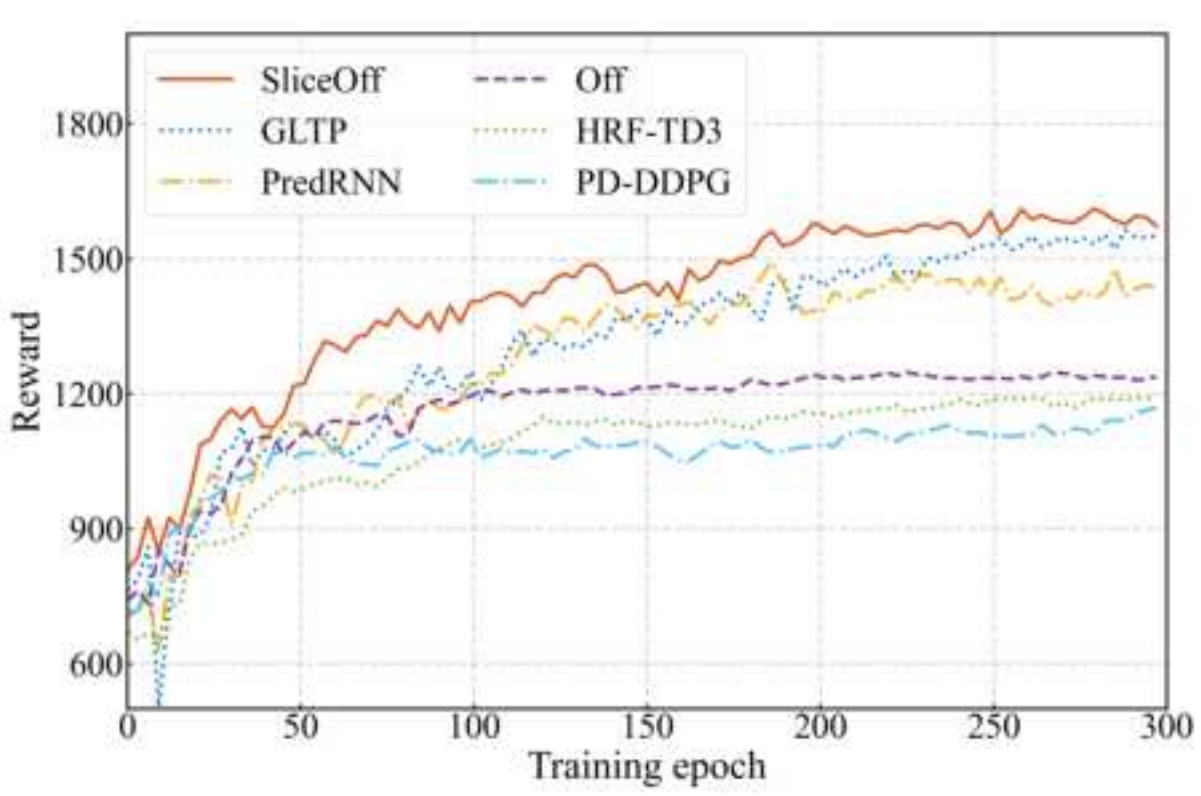

Fig. 4. Convergence of different methods.

**ESP profits of different methods.** We analyze the ESP costs, revenues, and profits by using different traffic prediction methods. As shown in Fig. 5, with dynamic slice adjustments, although the bandwidth/VM costs of the *SliceOff*, *GLTP*, and *PredRNN* are slightly higher than the *Off*, they can significantly enhance the revenues and profits. This is because static slice resource configurations cannot meet user demands as the traffic increases, causing the completion time of some tasks to exceed the maximum tolerable delay, thereby reducing revenues and profits. In the methods with dynamic slice adjustments, *SliceOff* achieves higher revenue with lower resource cost, which indicates that it has a more resource-efficient slice adjustment strategy. To explain this, we further compare the prediction accuracy of the three traffic prediction methods. Compared to the *GLTP* and *PredRNN*, the proposed *SliceOff* achieves higher prediction accuracy in terms of Mean-Square Error (MSE). This is because the *GLTP* and *PredRNN* simply adopt dilated convolutions and a gating mechanism to learn spatio-temporal traffic patterns. In contrast, the *SliceOff* introduces a self-attention mechanism that can better capture the sequence dependencies and solve the issues of gradient vanishing or explosion that may occur in *GLTP* and *PredRNN*, thereby promising higher prediction accuracy. The results demonstrate that more accurate traffic prediction can better support increasing resource efficiency and ESP profits.

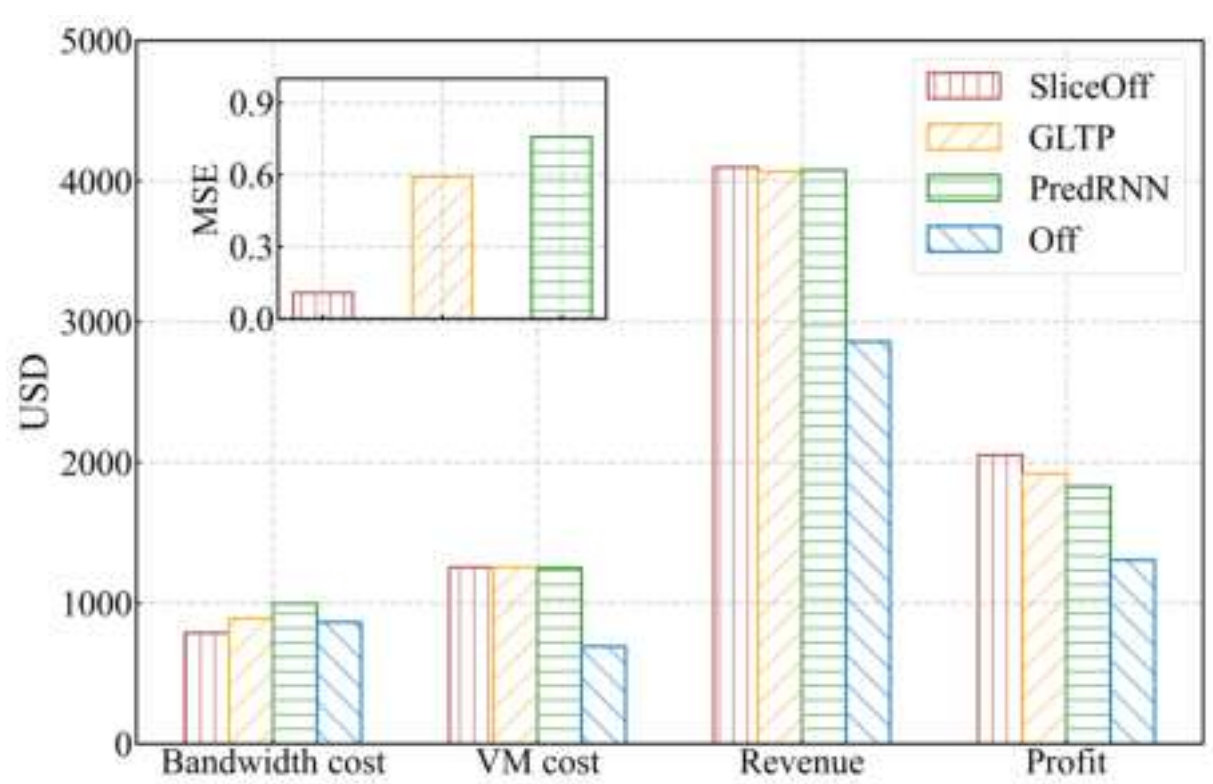

Fig. 5. ESP costs and profits of different methods.

**Task completion time of different methods.** We compare the task completion time of the *SliceOff* and other offloading methods with static slice resource configurations, consisting of uploading, queuing, and execution times. As illustrated in Fig. 6, the execution times of all methods are equal because all tasks are allocated with the same computing frequency. Compared to the other methods, the *HRF-TD3* and *PD-DDPG* exhibit much longer queuing and uploading times. This is because they cannot allocate appropriate bandwidth and VMs for the tasks, leading to long uploading latency and waiting queues on VMs. In contrast, the resource allocation implemented by *Off* is more efficient, thus reducing the task completion time. The uploading and queuing times of the *SliceOff* are significantly less than other methods. This is because the adaptive slice adjustments in *SliceOff* can dynamically adjust bandwidths and VMs according to traffic changes, thereby better meeting task demands. Meanwhile, we compare the average task completion times in different edge regions. The average task completion time varies between regions, but the completion times in all regions are lower than the maximum tolerable delay. The above results verify that the *SliceOff* can reasonably allocate resources to tasks based on task characteristics in different regions and thus achieve more efficient resource utilization.

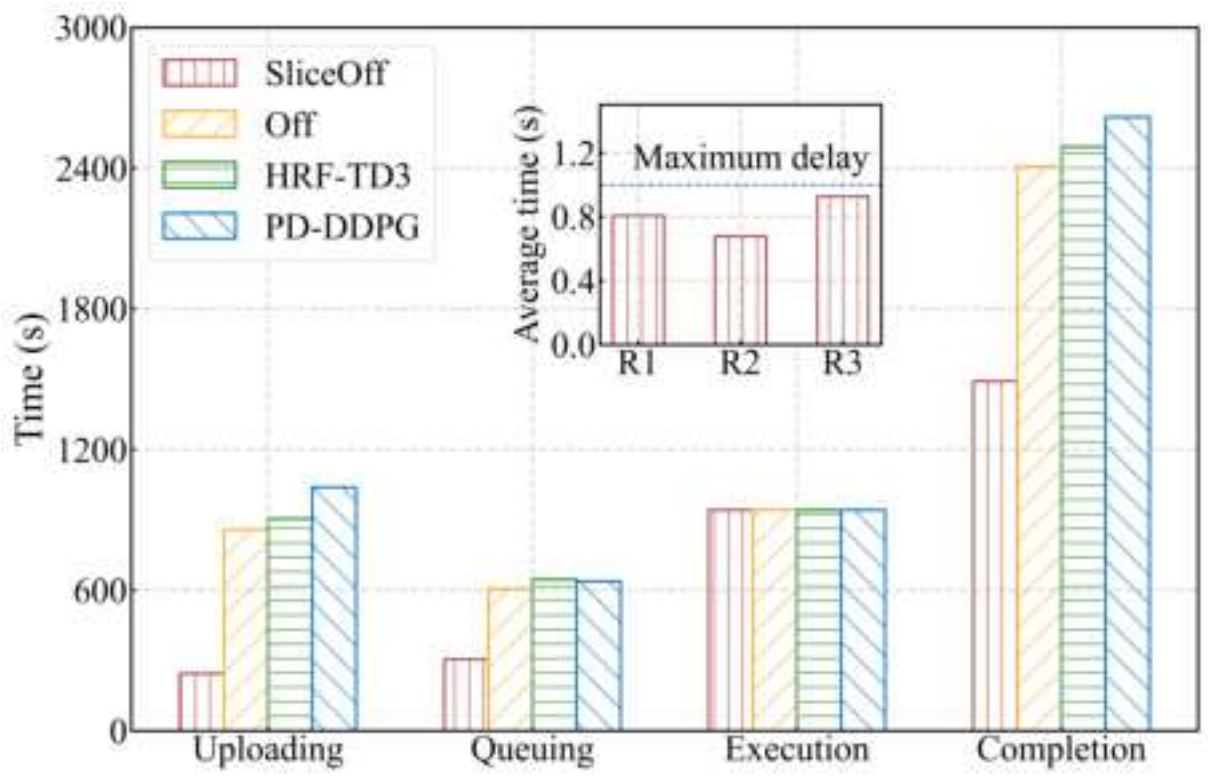

Fig. 6. Task completion time of different methods.

**Regional profits with various ratios of delay.** We analyze the impact of various ratios of delay $\omega$ on the ESP profits achieved by the *SliceOff* in different regions. As shown in Fig. 7, as the value of $\omega$ increases, the profits in three regions first increase and then decrease. The highest profit in each region is achieved when $\omega$ is 0.4, 0.3, and 0.2, respectively. This is because $\omega$ is used to control the tolerable communication and computing delays during slice adjustments, which also determines the ratio between the communication and computing resources rented by the ESP. For the region with high bandwidth costs (e.g., R1), the larger value of $\omega$ indicates that the available uploading time is longer, and thus the ESP can rent less bandwidth to meet task demands. For the region with high computing costs (e.g., R3), the smaller value of $\omega$ can reduce the VMs rented by the ESP. The results show that the proposed *SliceOff* can adaptively adjust slice resources according to the resource heterogeneity and task demands in different edge regions, demonstrating its practicality in performing slice adjustments in multi-edge systems.

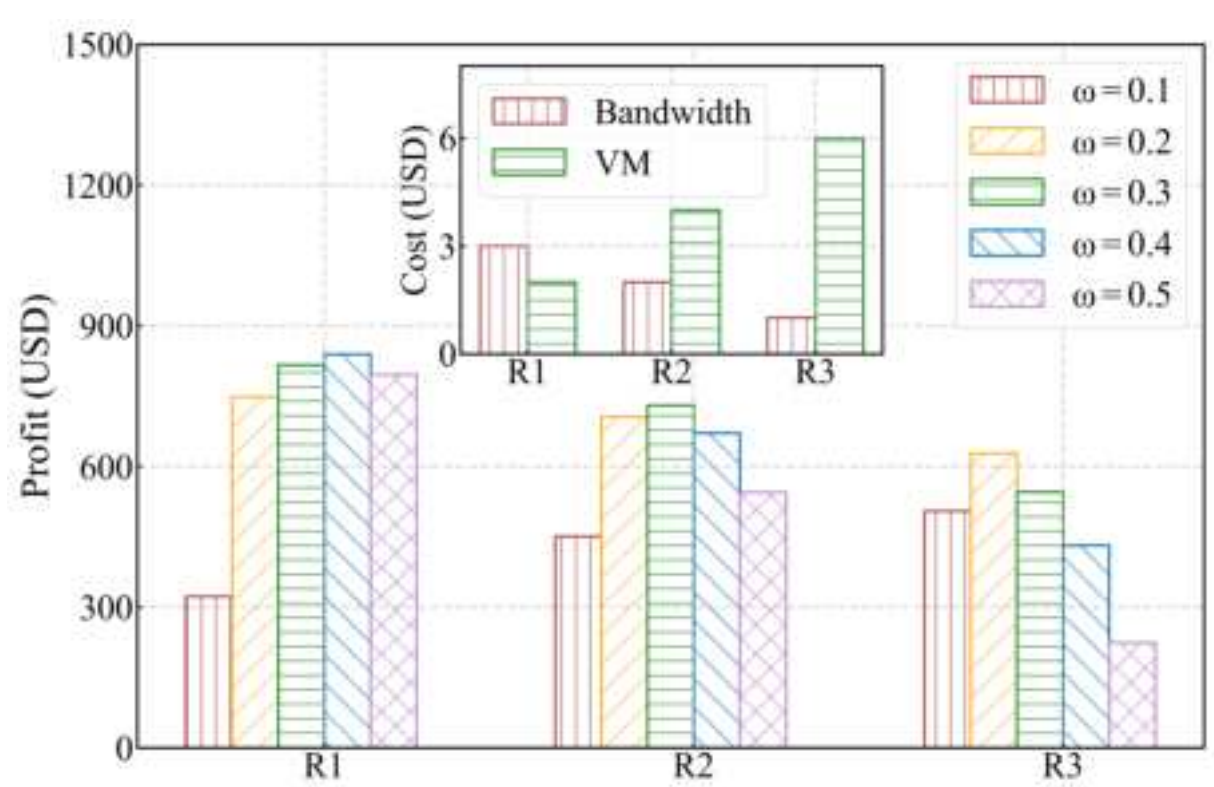

Fig. 7. Regional profits with various ratios of delay.

**RU with various multiples of traffic.** Fig. 8 presents the RU of different methods under the scenarios with various multiples of traffic. The RU grows as the traffic increases, where the RUs of the three methods with dynamic slice adjustments (i.e., the *SliceOff*, *GLTP*, and *PredRNN*) are higher than the other three methods without dynamic slice adjustments (i.e., the *Off*, *HRF-TD3*, and *PD-DDPG*). This is because only a few resources are allocated to offloaded tasks to meet user demands when the traffic is low, and thus the resources are not fully utilized. As the traffic increases, more resources are allocated and thus the RU rises. Compared to the *Off*, *HRF-TD3*, and *PD-DDPG* whose RUs change significantly as the traffic grows, the RUs of the other three methods are more stable. This is because the proposed slice adjustment can adaptively handle system load fluctuations caused by the dynamic traffic. The proposed *SliceOff* achieves the highest RU among all the methods because the designed self-attention-based traffic prediction method can accurately capture the spatio-temporal traffic changes, thereby improving the utilization of slice resources.

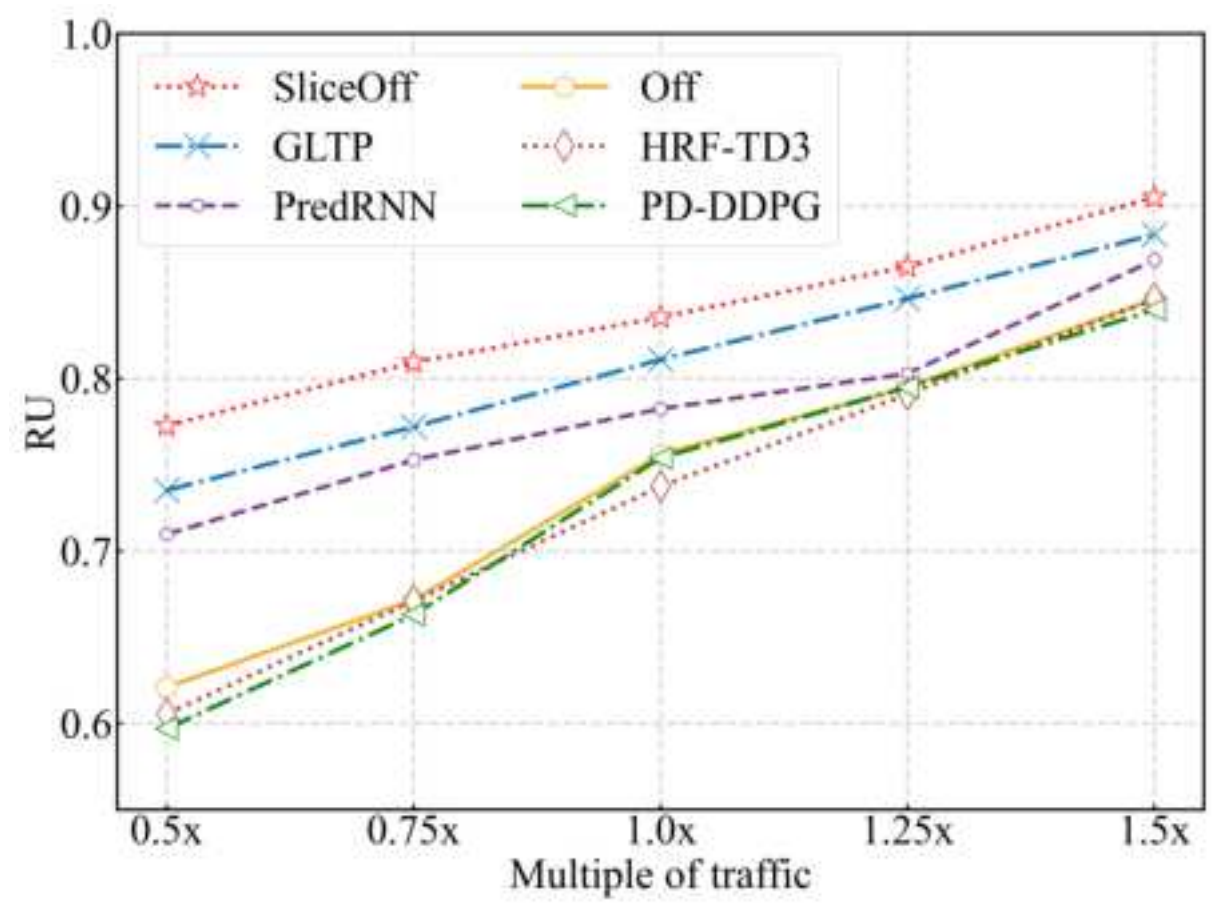

Fig. 8. RU with various multiples of traffic.

**DVR with various tolerable delays.** As illustrated in Fig. 9, we evaluate the DVR achieved by different methods under various task maximum tolerable delays. The DVR drops with the increase of the task maximum tolerable delay that decides the available time for completing a task. When the maximum tolerable delay is high, the DVR closes to 0, implying that the available resources can adequately satisfy the demands of processing tasks. Compared to other methods, the proposed *SliceOff* is able to complete more tasks within the maximum tolerable delay and thus reach a descending DVR. This is because the *SliceOff* design an improved DRL with dual-distillation, which improves the exploration efficiency and thus can adaptively make more reasonable decisions of network slicing and computation offloading based on dynamic user traffic and changeable system states.

### 5.3 Testbed Validation

**Settup of real-world testbed.** Based on hardware devices, we build a real-world testbed to further evaluate the practicality of the proposed *SliceOff*. As illustrated in Fig. 10, the testbed consists of three user devices (i.e., Raspberry 4B equipped with Broadcom BCM2711 SoC@1.5GHz) and three edge servers (i.e., Jetson TX2 equipped with Arm Cortex-A57 MPCore processor and 256-core NVIDIA Pascal GPU).

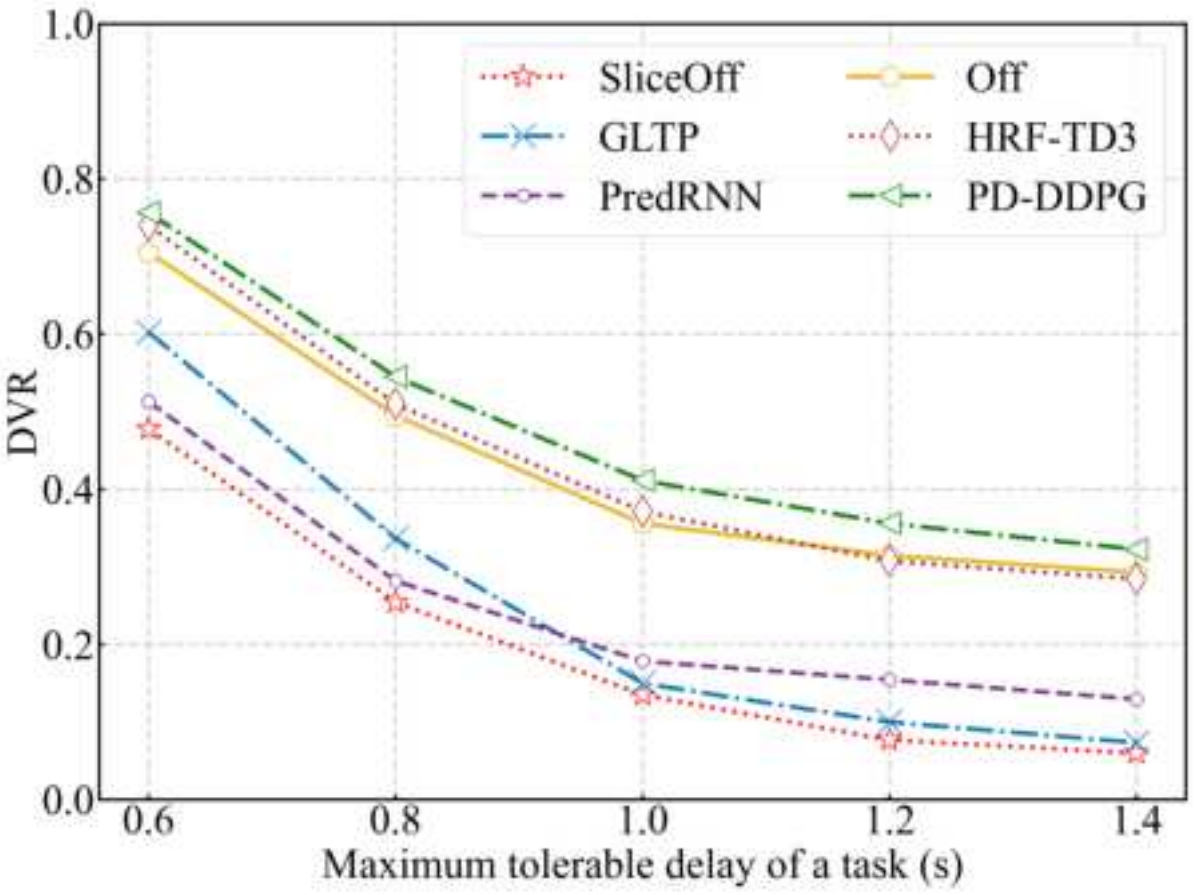

Fig. 9. DVR with various tolerable delays.

User devices are placed at different locations in the lab, which are treated as the three edge regions and generate tasks by following the same traffic as the simulation experiments. Three edge servers are placed together to provide computing resources for three edge regions in different time slots. All user devices and edge servers are connected through a 5GHz router, and the communication platform is built by the Flask framework. We regard the tasks of ResNet-based image classification as service instances for offloading. For different images, the time required for the edge server to perform inference is approximately 0.1$\sim$0.3 s. Specifically, we implement the *SliceOff* on an edge server and specify it as the ESP. After user devices generate image classification tasks, they first send offloading requests to the ESP. After receiving the offloading decisions from the ESP, the tasks can be uploaded to the target edge server and the results are returned after they are completed. Due to the different distances from the router, the time required for the three regions to upload the same task to the edge server is approximately 0.08 s, 0.15 s, and 0.2 s, respectively. For slice adjustments, we regard the number of edge servers as the slice units. When there are insufficient resources in a slice, tasks tend to be offloaded to an edge server. Otherwise, tasks can be offloaded to different edge servers.

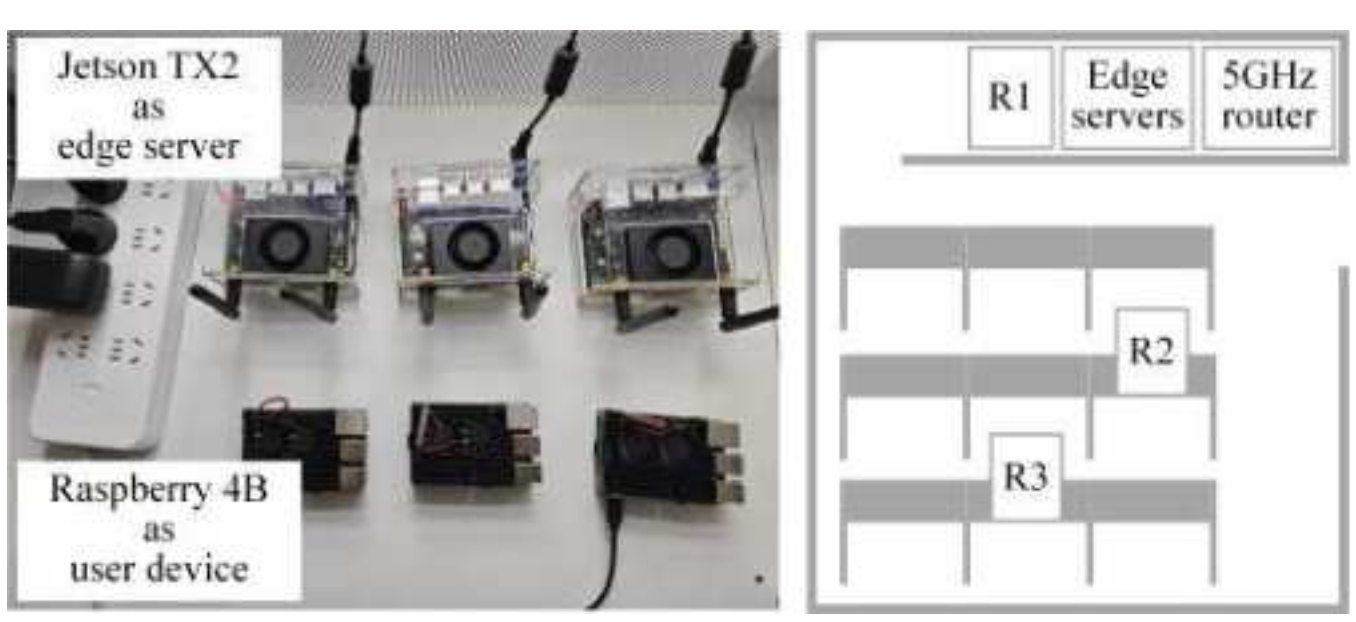

Fig. 10. Real-world testbed for evaluating the *SliceOff*.

**Task completion time in regions.** We evaluate the task completion time of various methods in different regions. The *SliceOff* can dynamically adjust resources (i.e., the number of available edge servers), while the *Off* and *MEC* offload tasks with the fixed number of edge servers. As depicted in Fig. 11, there are the maximum and minimum task completion times in R2 and R1, respectively. This is because there is the highest traffic in R2 and the distance from R3 to the router is the farthest, and thus they consume more task completion time than R1. This phenomenon indicates that user traffic and region locations affect the task completion time. In all regions, the task completion times achieved by the *SliceOff* and *Off* are less than the *MEC*. This is because the *SliceOff* and *Off* can effectively reduce the task waiting time by reasonably offloading tasks to edge servers according to the different distances and different task sizes. Notably, *SliceOff* achieves the lowest task completion time by dynamically adjusting the number of edge servers. The results demonstrate the superiority and practicality of the *SliceOff* in reducing the task completion time.

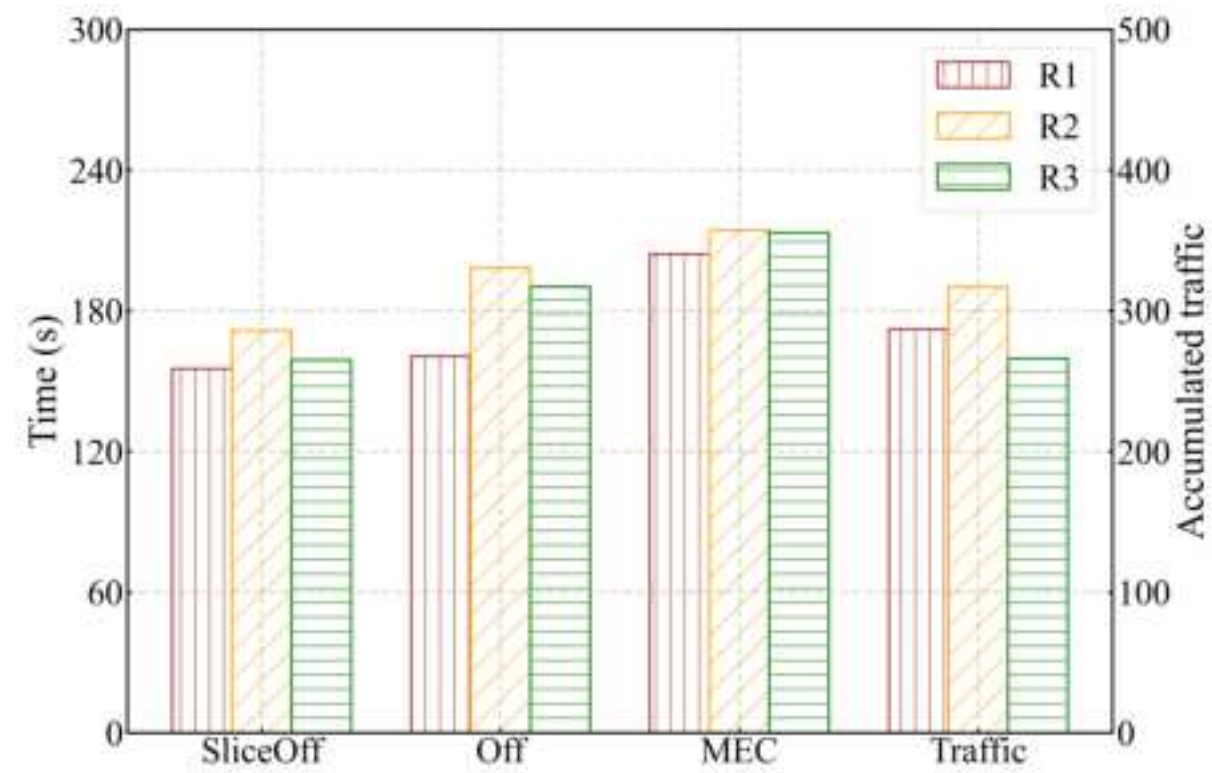

Fig. 11. Task completion time in different regions.

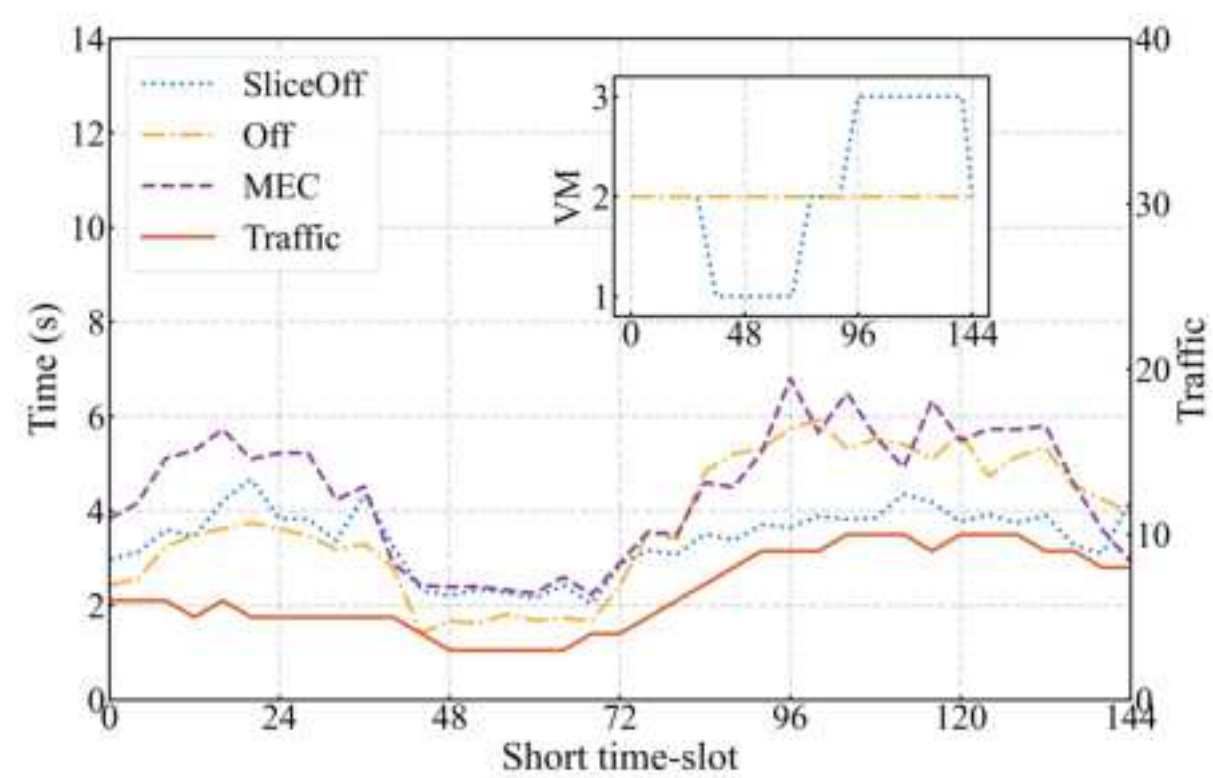

Fig. 12. Task completion time in different short time-slots.

**Task completion time in short time-slots.** We compare the total task completion time of three regions in different short time-slots. As shown in Fig. 12, the task completion times achieved by the three methods change with the traffic, indicating that the traffic fluctuations significantly impact the task completion time. Compared to the *Off* and *MEC*, the *SliceOff* significantly mitigates the impact of fluctuating traffic on system performance and overall achieves better performance. Specifically, the number of edge servers used by the *SliceOff* is adjusted between 1$\sim$3 according to the traffic while the number of edge servers used by the other two methods remains 2. When the traffic is low, the *SliceOff* reduces the available edge servers, which may cause a slightly higher task completion time than the *Off*. As the traffic rises, the *SliceOff* can effectively reduce the task completion time by dynamically increasing the edge servers, which is significantly lower than the *Off* and *MEC*. After testing, the server duration of the *SliceOff* is 298, and the server durations of the *Off* and *MEC* are both 288 ($144 \times 2$). The results verify that, by properly adjusting resources and then making offloading decisions, the *SliceOff* improves the efficiency of resource utilization while promising satisfying task completion time.

## 6 Conclusion

In this paper, we propose *SliceOff*, a novel resource-efficient offloading framework with traffic-cognitive network slicing for dynamic multi-edge systems. In *SliceOff*, we decouple the optimization problem from two perspectives: network slicing and computation offloading. For network slicing, we design a self-attention-based traffic prediction method and then combine linear programming and random rounding to achieve adaptive slice adjustments for enhancing the efficiency of resource utilization. For computation offloading, we develop an improved DRL method with a dual-distillation mechanism to improve the learning efficiency in huge offloading-decision spaces. Notably, we theoretically prove the effectiveness of the designed methods. With real-world datasets of user traffic, we validate the superiority of the *SliceOff* by extensive experiments. Compared to state-of-the-art methods, the *SliceOff* achieves better performance from the perspectives of ESP profits, task completion time, RU, and DVR under different scenarios. Further, the experiments on the real-world testbed validate the practicality of the *SliceOff*, which mitigates the performance imbalance caused by fluctuating traffic, thereby improving the task completion time and efficiency of resource utilization.